\documentclass{optica-article}

\journal{opticajournal} 

\articletype{Research Article}

\usepackage{lineno}

\usepackage{booktabs}
\usepackage{multirow}
\usepackage{array}
\usepackage[table]{xcolor}
\usepackage{makecell}

\begin{document}


\title{Scaleable LED-pumped Room-temperature Maser using a Multi-blade Optical Injector}

\author{
Mingyang Liu,\authormark{1,$\dagger$}
Zike Cheng,\authormark{2,$\dagger$}
Ziqiu Huang,\authormark{2}
Yifan Yu,\authormark{2}
Michael Newns,\authormark{2}
and Mark Oxborrow\authormark{2,*}
}

\address{
\authormark{1}Department of Physics, Imperial College London, Prince Consort Road, London, SW7 2AZ, United Kingdom\\
\authormark{2}Department of Materials and London Centre for Nanotechnology, Imperial College London, Prince Consort Road, London, SW7 2AZ, United Kingdom
}

\email{*m.oxborrow@imperial.ac.uk}

$\dagger$ These authors contributed equally to this work.


\begin{abstract*}
Though the performance of room-temperature masers has improved over the last decade, relatively little attention has been paid to the optics used to pump the maser's gain medium. In this work, we investigate a novel multi-blade optical ``injector'' that permits more effective and more scaleable pumping. 
The reported work encompasses an interdisciplinary mix of conceptualization, simulation, crystal growth, fabrication, and microwave engineering. Our gain medium is pentacene dissolved as a solid solution with \textit{para}-terphenyl (Pc:PTP) molecular crystal. We accurately determine this pentacene's molecular absorption cross-section as a function of wavelength. Ray-tracing is then used to assess how
different designs of waveguide inject light into the Pc:PTP crystal.
A multi-blade injector made of high-refractive-index glass (namely Ohara S-TIH6) is predicted to pump it more completely and uniformly than previous designs. Upon hand-fabricating such an injector and Bridgman-growing a crystal of 0.1\% Pc:PTP over it, an experimental maser oscillator using this combined injector-crystal assembly is demonstrated. The performance and scaleability of multiblade injection vis-\`{a}-vis alternative strategies is 
analyzed.
\end{abstract*}




\section{Introduction}
The maser, a device similar to a laser, can amplify microwave electromagnetic signals with exceptionally low-noise \cite{Reid2008Low-noiseNetwork}, making it indispensable for fields that require extreme sensitivity such as radio astronomy and deep-space communication \cite{poole1967electron}. However, traditional designs of masers, such as those based on ammonia\cite{gordon1958maser} or ruby\cite{kikuchi1959ruby}, require a high vacuum, or a strong applied magnetic field, or cryogenic conditions (or some combination of same) to operate. The development of room-temperature solid-state masers, such as the pentacene-doped \textit{para}-terphenyl (Pc:PTP) organic maser \cite{oxborrow2012room} and nitrogen-vacancy(NV)-center diamond maser \cite{breeze2018continuous}, has significantly reduced the size and cost of maser devices. While NV-diamond masers can operate continuously at room temperature, they still require a strong and uniform magnetic field (typically 430~mT) to achieve population inversion (at X-band) \cite{breeze2018continuous}. In contrast, the Pc:PTP maser can operate (at room temperature) in millisecond bursts without an external magnetic field, giving it greater immediate potential in portable, pulsed applications.
%
%
\begin{figure}[htbp]
  \centering
  \includegraphics[width = 0.9\textwidth]{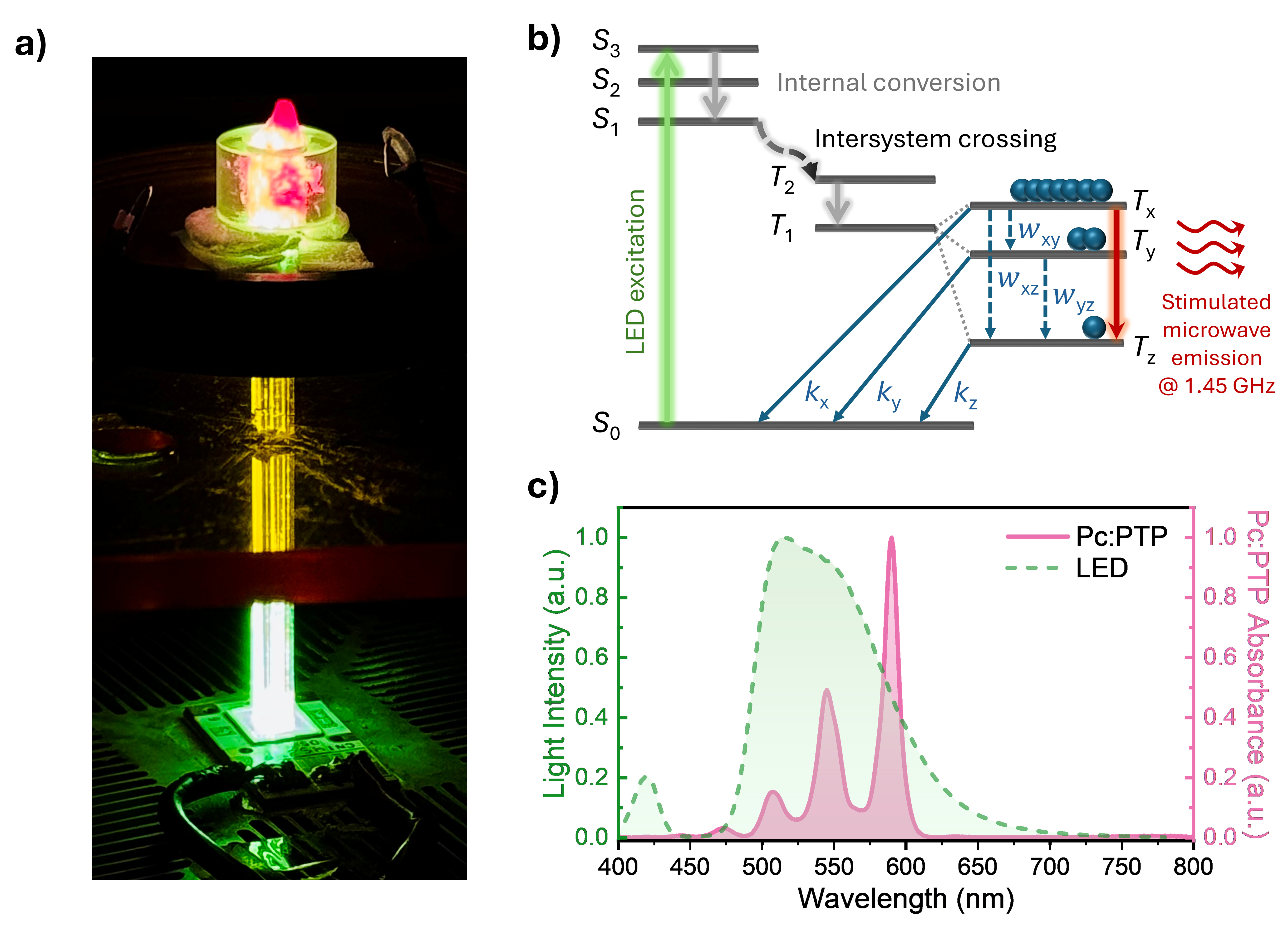}
  \caption{\textbf{Device set-up, simplified Jablonski diagram for Pc:PTP maser and UV-vis spectrum of Pc dissolved in PTP.} (a) A Pc:PTP crystal grown onto an ``aloe vera'' structured multi-blade waveguide under polychromatic LED-light illumination. (b)~LED light pumps the Pc molecules into the excited singlet $S_3$ state. These molecules then decay back to the $S_1$ state via (rapid) internal conversion. This is followed by intersystem crossing to the $T_2$ triplet state. The maser transition then occurs at 1.45 GHz from $T_x$ to $T_z$ within the $T_1$ state at ambient conditions. (c) UV-vis absorption spectrum of Pc:PTP (mauve-pink) and emission spectrum of pump LED (mint green).}
  \label{Jablonski}
\end{figure}
%
%
%
%
Modern telecommunications and measurement applications depend on low-noise amplifiers (LNAs) to boost the amplitude (thus power) of weak incoming signals such that they can be demodulated or otherwise detected by (digital) circuitry that is too noisy to perform the detection (without errors) directly. Without  sufficiently quiet and gainful LNAs, crucial messages get lost and critical measurements fail. When configured as an amplifier, the noisiness of a room-temperature maser\cite{Oxborrow2012MaserAssembly,Sherman2022Diamond-basedAmplifier}, as quantified by its noise temperature,
depends to a great extent on its so-called co-operativity, which (within the mean-field approximation) equals
%
%
%
%
%
%
\begin{equation}
\label{coop_mf}
\eta_{\mathrm{maser}}^\textrm{m.f.} = 
\left(
\frac{\mu_0 \gamma^2 \sigma^2  \Theta_\textrm{ISC}^\textrm{eff.}
T_1^\textrm{eff} T_2^{*} Q_\textrm{L}}
{2\pi f_{\mathrm{opt}}}
\right)
\cdot
\left(
\frac{
k_{\mathrm{opt}} P_{\mathrm{opt}} }
{V_{\mathrm{mode}}}
\right)
\end{equation}
where, in the numerator of the first quotient,  $\mu_0$ is the vacuum permeability, $\gamma$ is the electron gyromagnetic ratio (in units of Hz per tesla), $\sigma$ is the dimensionless interaction matrix element,
$\Theta^\textrm{eff}_\textrm{ISC}$ 
is the effective intersystem crossing efficiency, $T_1^\textrm{eff}$ is the effective relaxation time (of the emissive spin polarization), $T_2^*$ is the inhomogeneous dephasing time, $Q_\textrm{L}$ is the quality factor of the maser's (loaded) resonator. And, in this quotient's denominator, $f_\mathrm{opt} \equiv \omega_\mathrm{opt}/ (2 \pi)$ is the optical pump frequency. The numerator of the second quotient equals the total optical power absorbed by the maser crystal, $P_{\mathrm{abs}} \equiv k_{\mathrm{opt}} P_{\mathrm{opt}}$; its denominator is the maser mode's magnetic volume defined, within the mean-field approx., as
$V_{\mathrm{mode}} \equiv
\int \vert \textbf{B}(\vec{r}) \rvert^2 dV
 / \vert \overline{\textbf{B}(\vec{r})} \rvert^2$,
 where $\textbf{B}(\vec{r})$ is the mode's a.c.~magnetic flux density, the volume integral
 $\int ... \,\,dV$
 ranges over all spatial positions $\vec{r}$ where $\textbf{B}$ is non-zero (\textit{i.e.}, the inside of the microwave resonator's cavity) and $\overline{\textbf{B}(\vec{r})}$ is the mean flux density seen by the molecules whose spin-polarization drives the maser action.   
 
 In a real maser, the gain medium fills a region of space over which, in general, both the a.c.~magnetic flux density and the intensity of optical pumping vary. This requires the above expression for the co-operativity to be generalized to\cite{A.E.Siegman1964MicrowaveMasers}:
\begin{equation}
\label{coop_distrib}
\eta_{\mathrm{maser}} = 
\left(
\frac{\mu_0 \gamma^2 \Theta_\textrm{ISC}^\textrm{eff.}
T_1^\textrm{eff} T_2^{*} Q_\textrm{L}}
{\omega_{\mathrm{opt}}}
\right)
\cdot
\frac{
\sum_i \int p_{\mathrm{abs}}(\vec{r}) \{ \textbf{B}^*(\vec{r}) \cdot (\sigma \sigma^*)_i \cdot \textbf{B}(\vec{r})\} dV}
{\int \lvert \textbf{B}(\vec{r}) \rvert^2 dV}
\end{equation}
where $p_{\mathrm{abs}}(\vec{r})$ is the optical power absorbed by the maser crystal per unity volume --a scalar field, whose value depends, like the magnetic flux density, on the spatial position, $\vec{r}$;
and where $\sum_i$ sums over the different possible orientations (\textit{i.e.}, different magnetically inequivalent sites) of the emissive molecule in its crystal host, and
$(\sigma \sigma^*)_i$ is the dimensionless transition probability tensor\cite{A.E.Siegman1964MicrowaveMasers} for the $i$-th molecular
orientation. 

For the purposes of this paper, all of the parameters in the first factor (\textit{i.e.}, those inside the round bracket) in the above equation, and the design/geometry of the maser's microwave resonator itself (including its sample space), are regarded as givens. Towards boosting $\eta_{\mathrm{maser}}$ for an existing design of microwave resonator (in this case, a dielectric ring supporting a TE$_{01\delta}$ mode), we concentrate here on maximizing the volumetric integral over the available sample space in the numerator of the quotient on the right. Note here that Eq.~\ref{coop_distrib} assumes that the spin polarization generated by the pumping process is linear in the absorbed optical power and controlled by a single effective (relaxation) time constant $T_1^\textrm{eff}$\cite{Oxborrow2012MaserAssembly}. This will be true below the level of absorbed optical pump power density,
$p_\textrm{abs}^\textrm{sat}$ say, that causes saturation in the gain medium; this level itself will depend on the spatial concentration of spin-polarizable maser molecules in same and the time constant(s) controlling their spin dynamics. 

From equation~\ref{coop_distrib} we can see that, 
if the optical pump power is limited, it is most advantageous for it to be absorbed where the
a.c.~magnetic field $\textbf{B}(\vec{r})$ has a large amplitude [and favourably oriented with respect to the maser molecule's transition probability tensor $(\sigma \sigma^*)_i$]. In practice (with an apppropriately oriented crystal),  this means placing the maser crystal at the maser mode's magnetic antinode, with the mode's magnetic Purcell factor, namely~(ignoring constants) the fraction $Q_\textrm{L}/V_\text{mode}$, as large as possible. 
Even if saturation (in the spin-polarization mechanism) is not reached through such concentration/focussing, a practical limit is generally imposed by the need to prevent the maser crystal from loosing its performance (as a paramagnetic gain medium) or being irreversibly damaged through overheating. The avoidance of such ``hot-spots'' can be achieved by keeping $p_{\mathrm{abs}}(\vec{r})$ spread out, though the maser crystal's exact shape and thermal conductivity/diffusivity, as well as those of whatever materials surround it [each can potentially act as a
heat-spreaders and/or heat-sink/extractor, and/or (if liquid) as a coolant fluid] will also matter. Compared to the evolution of the internal combustion engine (into liquid- and air-cooled variants comprising multiple cylinders), the thermal  design of optically-pumped masers, in so far as how it might end up dictating the device's whole layout and geometry (as opposed to it entering as an  afterthought) is an underdeveloped area, though some recent progress in dual-using the metal (copper) that defines the maser's electromagnetic resonator as a heat sink/spreader for the
light-absorbing gain material (so as to avoid hot-spots), has been made\cite{fleury25}.

A non-zero value of 
$ p_{\mathrm{abs}}$ at a point in space within the gain medium requires a viable optical path between the maser's optical pump source and this point of absorption. This needed illumination can travel through air, or through the maser gain medium itself, or through some sufficiently transparent waveguiding structure. The attenuation of light through a linearly absorbing medium obeys the Beer--Lambert law,  with an absorption co-efficient of
$\alpha = - n \, \Xi$ nepers per unit length, where
$n$ is the spatial concentration of absorbing molecules and $\Xi$ is the molecular cross-section of an individual molecular. In, general, $\Xi$ will depend on the optical wavelength/frequency, as well as the directions of propagation and polarization (relative to the absorber's  molecular/crystal axes) of the light concerned.

A fundamental conflict/frustration in the design of an optically pumped maser is that $p_\textrm{abs}^\textrm{sat}$ and thereupon the absorbed power density $p_{\mathrm{abs}}(\vec{r})$ can be advantageously raised  by increasing the doping concentration $n$, yet do so will also increase the optical absorption co-efficient, $\alpha$, making it commensurately more difficult for light to travel deeply into the gain medium, so making it more difficult to achieve a spatially uniform  $p_{\mathrm{abs}}(\vec{r})$. An extreme example of this problem is that of ``sunburn", where an intense (air-pathed) pump laser hits the surface or a highly-doped maser crystal resulting in a thin surface layer of crystal being rapidly heated (and quite possibly thereupon burnt --irreversibly damaged), whilst the bulk of the crystal's interior is ``left in the dark'' and remains unpumped. We here present at least a partial solution to this problem.  Before describing what that is, we here briefly recognize that much can be generally learnt and borrowed from the (early) literature on optically pumped lasers\cite{siegman86}, whilst being mindful that, here, the (vacuum) wavelength of a microwave maser's output/oscillation is 5 orders of magnitude larger than that of its optical pump source, which is not the case with optically-pumped lasers.   
\begin{figure}[htbp]  
    \centering
    \includegraphics[width = 0.7\textwidth]{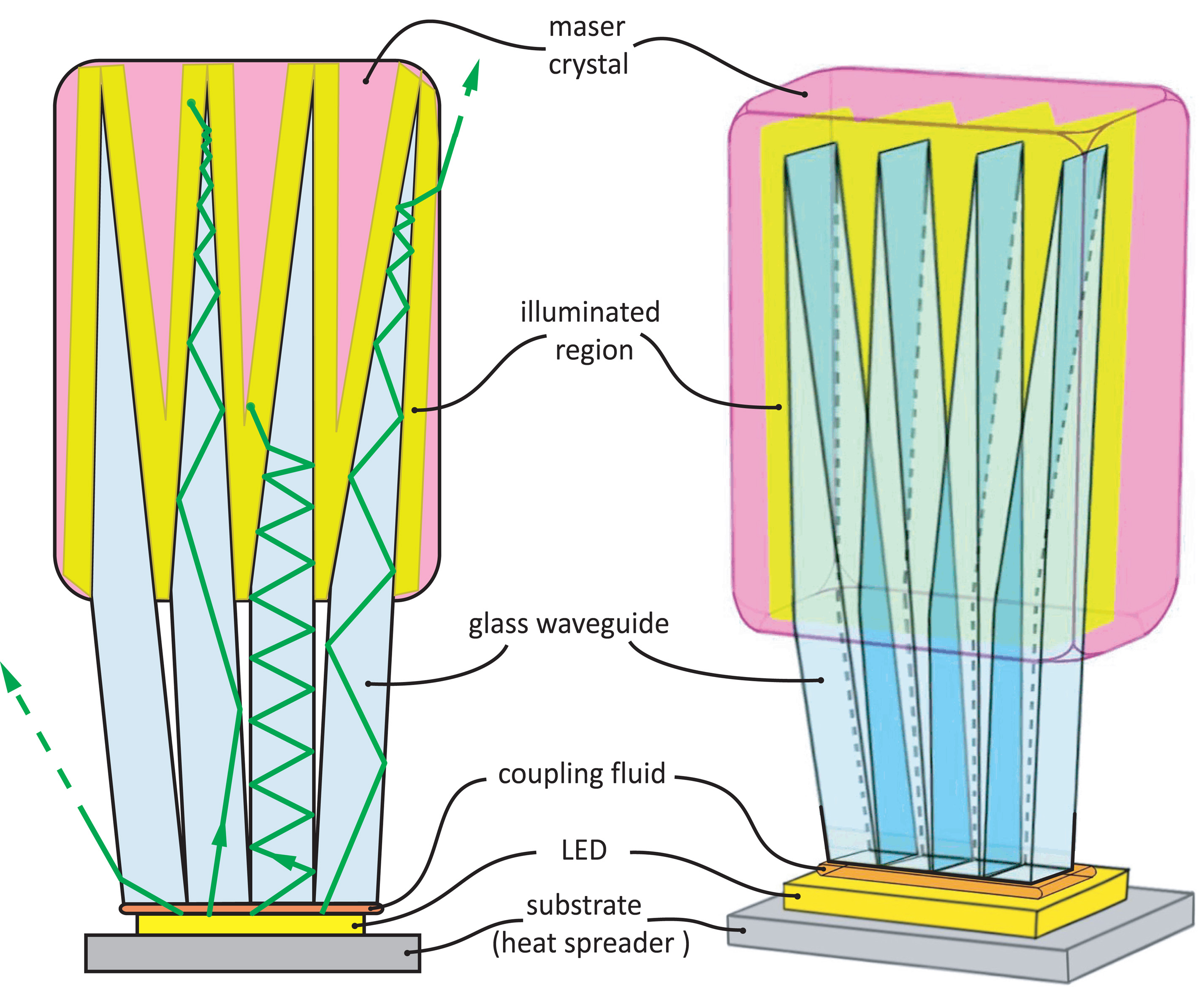}
    \caption{\textbf{Geometry of an ``aloe vera''-type invasive waveguide.} 
    The golden region identifies where the maser crystal is illuminated  by the LED. Note that the shanks of the waveguides in these two sketches are too short. In the left sketch (2D cross section), four different ray-trace scenarios are displayed (from left to right)\textbf{:} (i)~not captured by waveguide (angle of trajectory too acute)
    (ii)~oblique-angled propagation, (iii)~acute-angled propagation, (iv)~failure to be absorbed by maser crystal.  Both (ii) and (iii) end in absorption. }
    \label{ray-trace_explain}
\end{figure}
%
%
%

Accepting that the highest achievable co-operativities are obtainable through the use
of a heavily doped maser crystal whose optical penetration depth (for pump light) is
far smaller than its spatial dimensions, the only way to get light \textit{into} the bulk
of such a crystal is to replace a fraction of its volume by a network of optical waveguides.
The challenge  here is akin to providing a city with a transportation network or the working
cells of an organ (a liver, a lung or a kidney) with a supply of blood. Naturally,
as with civic planning or biological design, one tries to sacrifice as little active
volume (generating emissive gain) as possible to the construction of this supply network
whilst ensuring that it provides sufficient capacity and coverage. 
Assuming that this network has to be artificially constructed (one can dream of auto-intelligent self-assembly
processes, but ...), then it should as simple and easy-to-make as possible, without compromising
efficiency (in terms of capacity divided by volume occupied).

Two works by Wu~\textit{et al}\cite{Wu2020InvasiveCoupling,Wu2020Room-TemperatureConcentrator} and more recently Newns\cite{newns2025singleledpumpedroomtemperaturesolidstatemaser}
(see also Fig.~5 of Ref.~\cite{oxborrow2012room}) made first stabs (quite literally) at solving this design problem, introducing the concept of invasive optical pumping implemented
through spike- or wedged-shaped ends of optical waveguides sunk into the maser crystal. 
Analogous to the evolution of the internal combustion engine, from Otto's single cylinder to
banks of multiple cylinders (culminating in designs like the Napier Sabre, sporting 24 cylinders
in an H-block configuration), we here \textbf{bank up} a number of wedge-ended waveguides so as to more completely illuminate a 0.1\% doped crystal of pentacene-doped \textit{para}-terphenyl. Our basic design concept is shown in Fig.~\ref{ray-trace_explain}.    

\section{Experimental Work}
\subsection{Determination of Pentacene’s Optical Molecular Cross-section}
The rational, quantitative engineering of the maser's pump optical requires knowledge of the optical
properties of the materials through which light passes: coupling fluid, waveguide glass and the Pc:PTP itself. 
These properties are all known to sufficient accuracy, the one crucial exception being the optical absorption  coefficient of PC:PTP, which in turn, as already explained above,
depends on the product of the spatial concentration $n_\textrm{Pc}$ and optical cross-section of $\Xi_\textrm{Pc}$ of the pentacene molecules dissolved into the PTP. Surprisingly, we know of only one value of $\Xi_\textrm{Pc}$, namely $2 \times 10^{-17}$~cm$^2$, specifically for green 532~nm laser light propagating at normal incident w.r.t.~PTP's $a$-$b$ cleavage plane \cite{nelson1981laser}.
Given that our chosen light source emits over a range of wavelengths, we set about determining
$\Xi_\textrm{Pc}$ as a function of wavelength in this same direction. 
A set of optical samples of pentacene-doped \textit{para}-terphenyl at 3 different doping concentrations, whose plane-parallel optical surfaces (`windows') followed the $a$-$b$ cleavage plane (to within a tolerance of a few degrees) were prepared using the method detailed in Appendix I. The thickness of each (delicate) sample was determined (upon carefully standing the sample on its side against a graticule) through optical microscopy.   

\begin{figure}
    \centering
    \includegraphics[width = 0.8\textwidth]{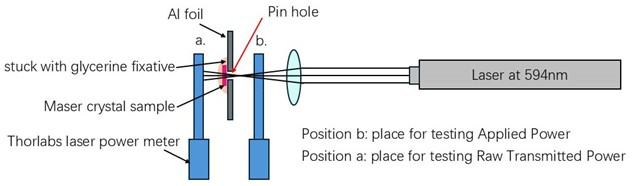}
    \caption{\textbf{Optical set-up for determining absorption through Pc:PTP sample.}}
    \label{fig:Maser_experiment_optical}
\end{figure}

\begin{figure}[b]
  \centering
  \includegraphics[width = 0.6\textwidth]{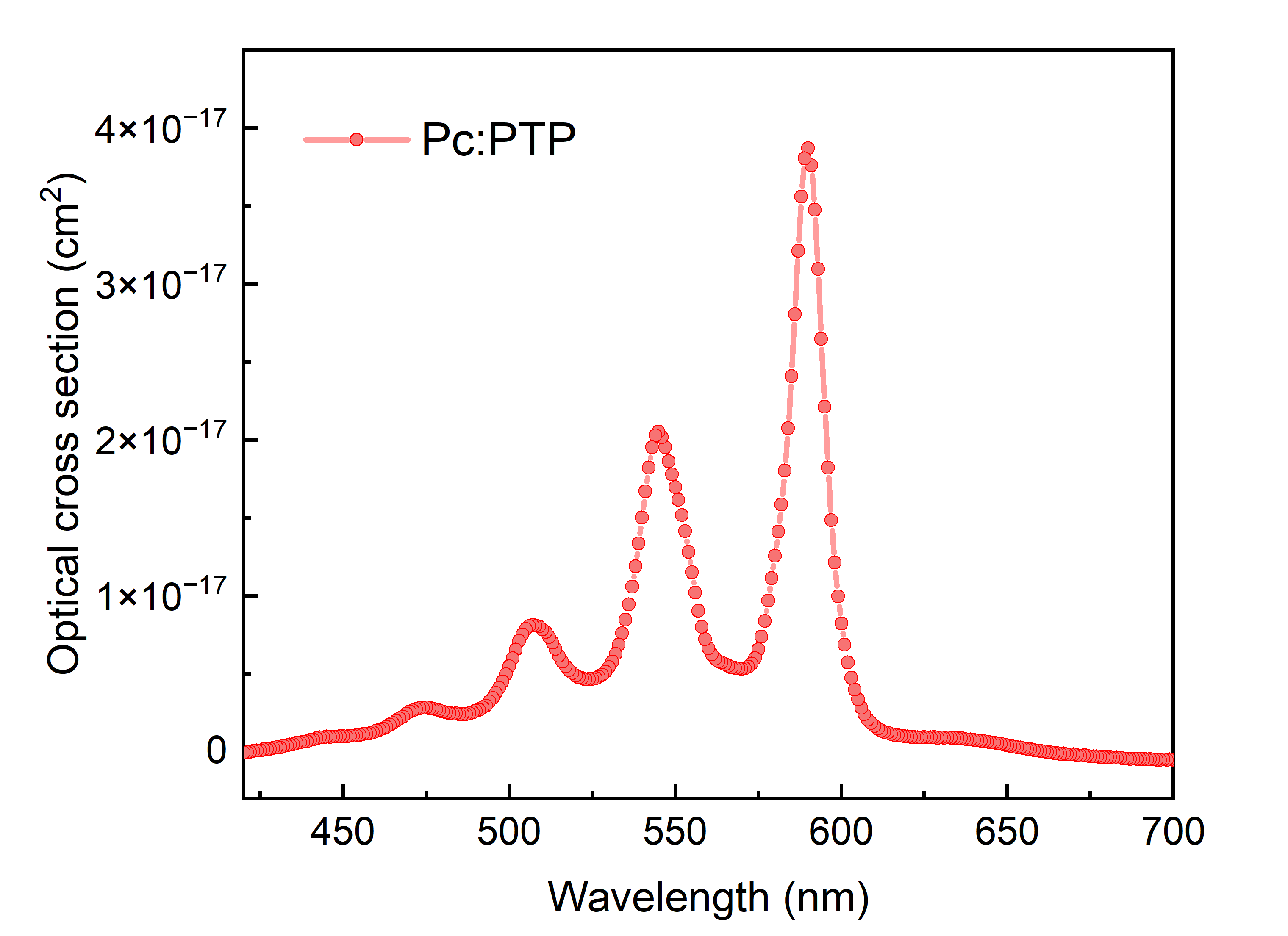}
  \caption{\textbf{Optical absorption cross-section of 0.1\% pentacene dissolved in solid \textit{para}-terphenyl.} For unpolarized light traveling perpendicularly to the crystal's primary growth facet ($a$-$b$ cleavage plane).}
  \label{mol_cross}
\end{figure}
\begin{table}[htbp]
    \caption{\textbf{Transmittance of the laser beam through Pc:PTP maser crystals at three different concentrations, together with the inferred optical parameters.}}
    \label{tab:transmittance}
    \centering
\resizebox{\textwidth}{!}{%
\begin{tabular}{cccccccc}
\textbf{Pc Conc.} &
\textbf{Power} / mW &
\textbf{Raw Trans. Power} / mW &
\textbf{Trans. Ratio} &
\textbf{Real Trans.} &
\textbf{t} / mm &
\textbf{$\alpha$} / mm$^{-1}$ &
\textbf{$\Xi_\textrm{Pc}$ at 596 nm} / cm$^{2}$ \\
\hline
0.1\%  & 4.40 & 1.04 & 0.2364 & 0.2660 & 0.133 & 9.959  & $2.17 \times 10^{-17}$ \\
0.05\% & 4.41 & 1.70 & 0.3854 & 0.4339 & 0.152 & 2.749  & $1.37 \times 10^{-17}$ \\
0.01\% & 4.42 & 3.38 & 0.7647 & 0.8607 & 0.171 & 0.8207 & $2.46 \times 10^{-17}$ \\
\hline
\end{tabular}
}
\end{table}

A narrow beam of yellow-orange 596 nm light, generated by a special type of He-Ne laser, was directed through the optical windows of each crystal samples. The power of this beam was just a few mW. A convex lens was used to focus the beam down to a narrow “waist” where the sample was inserted; see Fig.~\ref{fig:Maser_experiment_optical} To prevent optical “spill-over” (\textit{i.e.}, light passing by and not through the crystal) into the optical detector, the sample is mounted using petroleum jelly (purely as a fixative) over a pinhole in a piece of aluminum foil. The power of the laser beam before and after transmission was measured using the laser power meter. The ratio of these two measured powers provides the raw transmittance. This value is then modified using Fresnel's equations to compensate for reflections at the two air: \textit{para}-terphenyl interfaces to provide the true transmittance through the crystal and thereupon (knowing the sample's thickness) the
Beer--Lambert absorption coefficient $\alpha$; see Table~\ref{tab:transmittance}.
The same optical samples (fixed over pinholes) were inserted into a Cary 5000 UV-Vis-NIR spectrophotometer; the resultant optical absorption spectra enabled the molecular cross-section as a function of (vacuum) wavelength, namely $\Xi_\textrm{Pc}(\lambda)$, as shown in Fig.~\ref{mol_cross}, to be independently determined. 
We acknowledge the discrepancy between the value displayed in this figure at $\lambda$ = 596~nm, and the corresponding (lower) values displayed in the right-most column of Table~\ref{tab:transmittance}. A lack of spatial homogeneity (non-uniform doping), or inadvertent misorientation during measurement, in one or several of the optical samples, could easily explain it.

\subsection{Fabrication of Multiblade Injector}

\begin{figure*}[]
 \centering
  \includegraphics[width=0.9\textwidth]{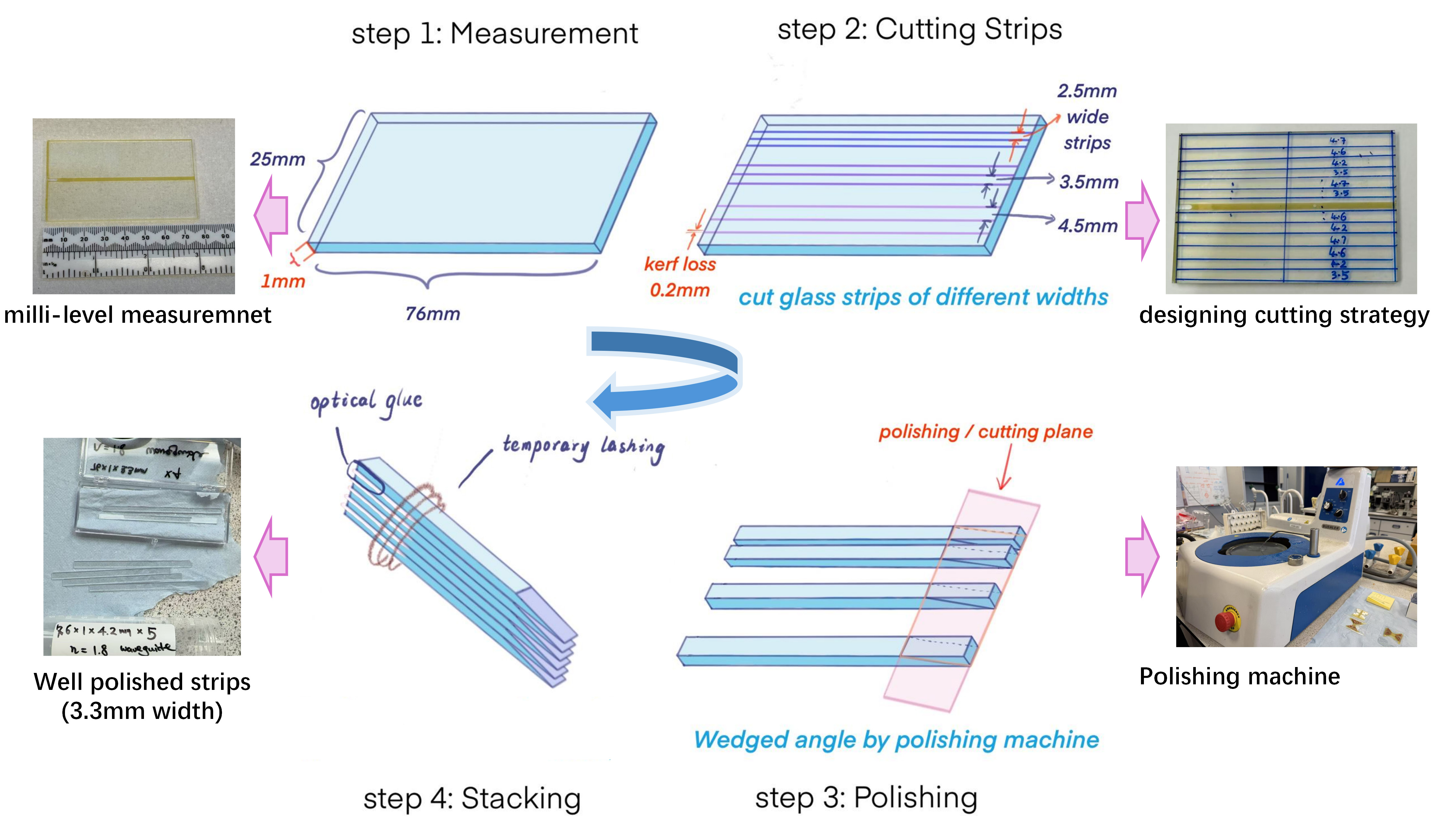}
  \caption{\textbf{Workflow for preparing multi-blade waveguides for maser experiments.} Starting from a 25 mm × 76 mm glass slide (step 1), the substrate is measured and sectioned into narrow strips of defined widths (2.5–4.5 mm), accounting for kerf loss during cutting (step 2). The resulting strips resemble uniform ribbons. Each strip is then mounted in the polishing machine and processed to achieve a controlled wedge angle on one face (step 3). In step 4, multiple polished strips are aligned and temporarily stacked by using optical glue, lashing and clips to facilitate handling and further processing.}
  \label{wave_guide_make}
\end{figure*}

\begin{figure}[]
  \centering
  \includegraphics[width=0.45\textwidth]{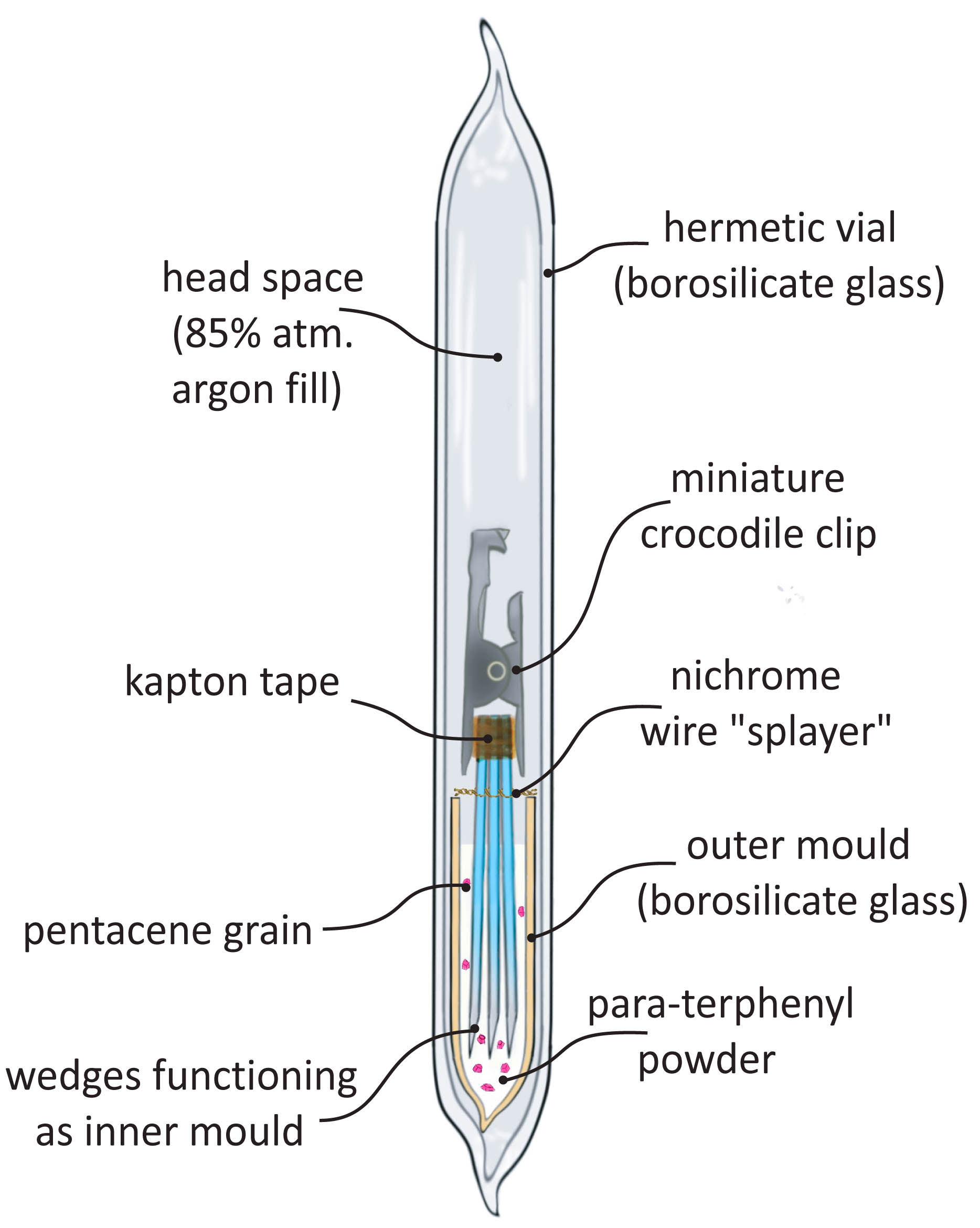}
  \caption{\textbf{Loaded Bridgman growth vial (before dropping through oven).}}
  \label{growth_mould}
\end{figure}

A plate of Ohara S-TIH6 (n = 1.8) glass was cut into slices using a wafer saw and polished with a metallographic grinder.
Waveguides were cleaned sequentially in \textit{p}-xylene, acetone, and isopropanol with an ultrasonic bath to remove organic residues and ensure a clean, dry surface for subsequent processing. The waveguides were then assembled and placed into  5.9 mm inner-diameter borosilicate vial. After several evacuation and back-fills with argon gas the vial was sealed (``tipped'') with a flame; see
Fig.~\ref{growth_mould}. The vial was then dropped through a vertical tube furnace over two days.  
Any extra PTP that had solidified on the waveguide was then removed with dental tools after immersion in toluene for 1 day.

\subsection{Demonstration of Masing}
\begin{figure}
    \centering
    \includegraphics[width = 0.9\textwidth]{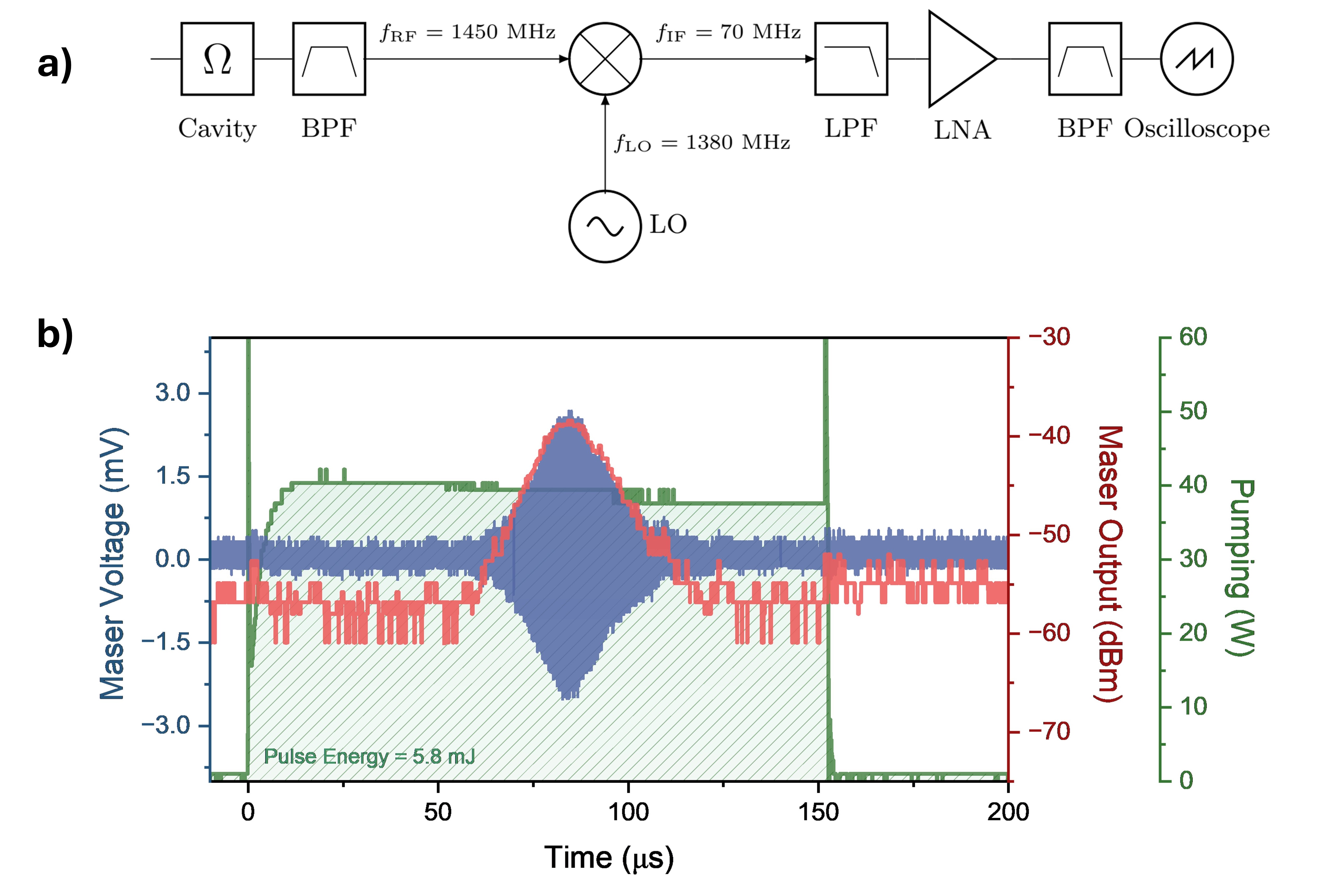}
    \caption{\textbf{Experimental masing set-up and spectrum.} (a) Schematic of the heterodyne detection chain used to monitor the maser emission. (b) Time-domain maser signal.}
    \label{fig:Maser_experiment}
\end{figure}

A maser was then constructed around the Pc:PTP maser crystal integrated with its Ohara multi-blade injector: The injector was inserted into a ring of strontium titanate surrounded by a copper housing with adjustable ceiling\cite{Breeze2015EnhancedMasers}; see Fig.~\ref{Jablonski}(a). This geometry supports a frequency-tunable high-Q TE$_{01\delta}$ mode at around 1.45@GHz.
The waveguide was fed at its bottom end by an OSRAM OSTAR LE A P3MQ LED \cite{osmar} producing yellow-green 565-575~nm light.
This LED's outputting surface is a 3.2~mm $\times$ 2.6~mm rectangle, which the banked-together waveguides just cover.
The gap between them and the LED is filled by a drop of `Santovac 5' polyphenyl ether diffusion-pump oil.
Upon applying 21~V for 150~$\mu$s across the LED through the use of a MOSFET switch,
a robust maser burst, delayed by approximately 60~$\mu$s from the onset of pumping
is robustly observed. The maser output was measured using a single-stage superheterodyne receiver incorporating a 70-MHz IF strip
as shown in Fig.~\ref{fig:Maser_experiment}(a). The mixed-down signal was recorded using a RIGOL DS1104 oscilloscope (100 MHz 1~GSa~s$^{-1}$); see Fig.~\ref{fig:Maser_experiment}(b).

\section{Optical Simulation, Design and Optimization}

\subsection{1D Model for Optical Pumping Regime Identification}
In this work, the one-dimensional model is used to delineate the optical pumping regimes in which a single effective absorption coefficient provides a valid first-order description. This enables a consistent and physically grounded geometric analysis of different injector designs in three dimensions. While the model explicitly resolves the populations of the ground state and individual triplet sublevels, its role in the present work is to identify the optical pumping regimes relevant to geometric injector optimization.

Optical pumping of a Pc:PTP maser crystal is intrinsically non-linear due to the accumulation of population in long-lived triplet states ($T_x, T_y, T_z$), which depletes the singlet ground state $S_0$ and suppresses further optical absorption. As a result, the local absorption coefficient becomes intensity dependent, and the spatial attenuation of pump light can deviate from simple Beer--Lambert behaviour at sufficiently high pump intensities.

In the present work, three-dimensional ray-tracing simulations are used to compare different optical injector geometries. These simulations necessarily assume a linear Beer--Lambert absorption model characterized by a single absorption coefficient. A necessary preliminary step is therefore to determine, using a more complete microscopic description, the range of pumping intensities over which such an effective absorption coefficient provides a reasonable first-order description of the longitudinal absorption behaviour. To this end, we employ a one-dimensional coupled optical--spin-dynamics model to quantify how bleaching modifies the absorption profile and to extract an intensity-dependent effective penetration depth.

The model resolves optical propagation and molecular population dynamics along a representative longitudinal axis $x$ of the crystal. It captures bleaching and longitudinal saturation effects arising from ground-state depletion but does not attempt to describe transverse redistribution of optical power. Spatial redistribution due to injector geometry is instead treated explicitly in our three-dimensional ray-tracing simulations presented further below.  The role of the present 1D model is therefore to provide an effective longitudinal absorption description that can be used consistently as an input to the geometric analysis.

The molecular system is described using a four-level scheme comprising the singlet ground state ($S_0$) and the three triplet sub-levels ($T_x$, $T_y$, $T_z$). The first excited singlet state ($S_1$) is treated within a quasi-steady-state approximation (QSSA), which is valid throughout the pumping intensity range considered here and was verified by comparison with a full five-level model. Under this approximation, the stimulated absorption rate $W$ directly partitions population flow from $S_0$ into the competing fluorescence and intersystem crossing (ISC) pathways, allowing bleaching-induced suppression of absorption to be captured self-consistently.

At a given time $t$ and position $x$, the population dynamics are governed by the following rate equations:
\begin{subequations}
\begin{align}
\dot{N}_x &= R_x N_1 - k_{x} N_x 
       + w_{xy} (N_y - N_x) 
       + w_{xz} (N_z - N_x) \notag \\
\dot{N}_y &= R_y N_1 - k_{y} N_y 
       + w_{xy} (N_x - N_y) 
       + w_{yz} (N_z - N_y) \notag \\
\dot{N}_z &= R_z N_1 - k_{z} N_z 
       + w_{xz} (N_x - N_z) 
       + w_{yz} (N_y - N_z) \notag \\
\dot {N}_0 &+ \dot{N}_x + \dot{N}_y + \dot{N}_z = 0
\label{eq:rate}
\end{align}
\end{subequations}
subject to the initial condition:
\begin{equation}
{N}_0 (t=0) = N_\textrm{total}
\end{equation}
where $N_0$, $N_x$, $N_y$ and $N_z$ are the population densities of $S_0$, $T_x$, $T_y$ and $T_z$ respectively; $N_\textrm{total}$ is the total population density of all $k_x$, $k_y$ and $k_z$ are the zero-field reverse-ISC rates from $T_x$, $T_y$ and $T_z$ to $S_0$ respectively; and $w_{xy}$, $w_{xz}$ and $w_{yz}$ are the spin-lattice relaxation rates between triplet sublevels (see Figure \ref{Jablonski}a). $R_x$, $R_y$ and $R_z$ are the optical pumping rates from $S_0$ to $T_x$, $T_y$ and $T_z$ respectively, given by:
\begin{equation}
R_i = P_i \phi_\textrm{ISC} W
\end{equation}
where $P_i$ is initial relative population ratio of $T_i$ sublevel following intersystem crossing (ISC); $\phi_\textrm{ISC}$ is the effective ISC yield, given by:
\begin{equation}
\phi_\textrm{ISC} = \frac{k_\textrm{ISC}}{k_\textrm{ISC} + k_{S_1\to S_0} + W}
\end{equation}
where $k_\textrm{ISC}$ is the zero-field total ISC rate from $S_1$ to the triplet states; $k_{S_1\to S_0}$ is the total depopulation rate (here we approximate it as the fluorescent rate) from $S_1$ to $S_0$; and $W$ is the stimulated absorption rate (optical pumping rate / photoexcitation rate) from $S_0$ to $S_1$, given by:
\begin{equation}
W = \int_{0}^{\infty} \frac{I(\omega) \sigma(\omega)}{\hbar\omega} d\omega
\end{equation}
where $I(\omega)$ is the local light intensity, $\sigma(\omega)$ is the absorption cross section, at angular frequency $\omega$; $\hbar$ is the reduced Planck constant.
The model assumes Kasha's rule, wherein any excitation to higher singlet states ($S_n, n>1$) undergoes rapid and 100\% efficient internal conversion to the lowest excited singlet state ($S_1$). This stimulated absorption rate $W$, then feeds the competing total depopulation rate ($k_{S_1\to S_0}$) and intersystem crossing ($k_\textrm{ISC}$) pathways. 

\begin{table}[b]
\caption{\textbf{Rates parameters}
The units of each transition rate $k$ are $10^{4}$ s$^{-1}$}
\label{tab:parameters}
\centering
\renewcommand{\arraystretch}{1.3}

\begin{tabular}{c c c c c c}
\toprule

\rowcolor{gray!25}
\multicolumn{3}{c}{ISC rate, $k_{\mathrm{ISC}}$} &
\multicolumn{3}{c}{Fluorescent rate, $k_{S_1\to S_0}$} \\
\midrule

\multicolumn{3}{c}{$1.82\times10^{5}$} &
\multicolumn{3}{c}{$1.1\times10^{5}$}%
\makebox[0pt][l]{\hspace{1em}ref.~\cite{takeda2002zero}} \\
\midrule

\rowcolor{gray!25}
\multicolumn{6}{c}{Relative population rates, $P_x:P_y:P_z$} \\
\midrule

\multicolumn{6}{c}{0.732:0.159:0.109}
\makebox[0pt][l]{\hspace{1em}ref.~\cite{mann2025chemically}} \\
\midrule

\rowcolor{gray!25}
$k_x$ & $k_y$ & $k_z$ &
$w_{xz}$ & $w_{yz}$ & $w_{xy}$\\
\midrule

\multicolumn{1}{c}{2.37} &
\multicolumn{1}{c}{1.20} &
\multicolumn{1}{c}{0.45} &
\multicolumn{1}{c}{0.15} &
\multicolumn{1}{c}{2.19} &
\multicolumn{1}{c}{1.17}
\makebox[0pt][l]{\hspace{1em}ref.~\cite{mann2025chemically}} \\

\bottomrule
\end{tabular}

\end{table}


The light intensity propagation is described by the following equation:
\begin{equation}
\frac{\partial I(\omega)}{\partial x} = -\sigma(\omega) (N_0 - N_1) I(\omega) \approx -\sigma(\omega) N_0 I(\omega)
\label{eq:lightprop}
\end{equation}
where $x$ is the propagation distance in the crystal. $N_1$ is assumed to be negligible compared to $N_0$ in our QSSA model.
%
%
The key simulation parameters used in this model are summarized in Table~\ref{tab:parameters}.
%
%

\begin{figure*}[htbp]
  \centering
  \includegraphics[width=0.8\textwidth]{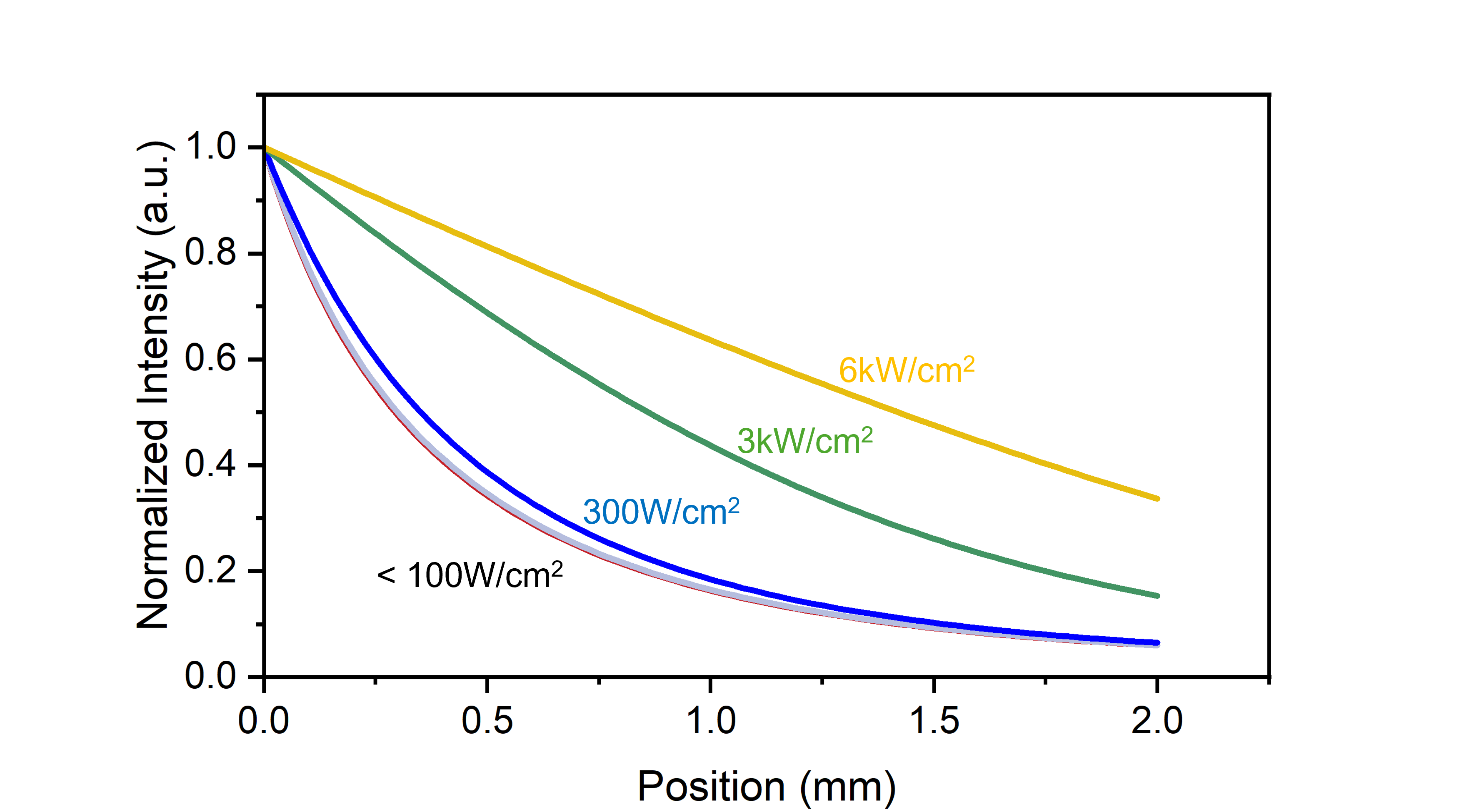}
  \caption{\textbf{Steady state absorption curves with different input intensity (for our LED's polychromatic output spectrum)}}
  \label{absorption_curve}
\end{figure*}

Quasi-steady-state simulations were performed for a 2~mm-thick Pc:PTP crystal with a pentacene concentration of 0.1\%. At low pump intensities ($I \lesssim 100~\mathrm{W\,cm^{-2}}$), corresponding to the present LED-invasive-pumped experiments (waveguide output intensity$I \approx 30~\mathrm{W\,cm^{-2}}$), the absorption profile is well described by an exponential Beer--Lambert law with a penetration depth of approximately 0.47~mm. In this regime, the fraction of molecules occupying triplet states remains below $\sim$1\% within the first millimetre of propagation.
At higher pump intensities ($I \gtrsim 200~\mathrm{W\,cm^{-2}}$), ground-state depletion begins to become significant, and the absorption profile deviates from a simple exponential form. As bleaching suppresses absorption along the propagation direction, the effective penetration depth increases. At $I \approx 3000~\mathrm{W\,cm^{-2}}$, the penetration depth extracted from an exponential fit increases to approximately 1.18~mm, reflecting a substantial reduction in the average absorption strength due to triplet accumulation. The resulting absorption profiles for different pump intensities are summarized in Fig.~\ref{absorption_curve}.

While the absorption becomes increasingly non-exponential at high intensities, the effective penetration depth provides a useful first-order scalar measure of the average longitudinal attenuation experienced by pump light. This effective description enables a consistent comparison of different injector geometries within the three-dimensional ray-tracing framework, where absorption is necessarily treated using a linear Beer--Lambert model with a single absorption coefficient.
Within this framework, differences observed in the three-dimensional simulations arise purely from geometric light delivery and spatial redistribution of optical power, rather than from differences in intrinsic material response. 

\subsection{3D Geometric Analysis of Optical Power Redistribution}
\label{subsec:shape_optimisation}

Throughout the following analysis, we do not aim to maximize purely geometric or normalized uniformity. Instead, we evaluate spatial uniformity under the constraint of finite and identical effective absorption, since a perfectly uniform but weakly absorbing design would contribute negligibly to maser cooperativity. All uniformity metrics are therefore interpreted in an engineering sense: they quantify how effectively a given absorbed optical power is distributed over the crystal volume.

The geometric emission characteristics of the LED pumping source play a central role in determining how optical power is delivered and redistributed within the maser crystal. Unlike lasers, which provide highly collimated beams, LEDs emit strongly divergent radiation that approximately follows a Lambertian angular distribution in air. Upon refraction into the waveguide, this distribution is truncated by the critical angle and further modified by Fresnel transmission, resulting in a highly non-uniform angular intensity profile inside the waveguide.

To guide the analysis, the waveguide is conceptually divided into two regions: the \textbf{rod part}, located in air, and the \textbf{blade part}, embedded in the Pc:PTP crystal (refractive index $n \approx 1.63$). For the rod part, the waveguide refractive index must be sufficiently high to ensure total internal reflection (TIR) for all rays entering from LED, preserving input angular distribution along the propagation axis. For the blade part, a refractive index lower than that of the surrounding crystal would result in immediate light leakage, while an excessively high index would cause over-propagation and poor optical filling. 

These competing requirements define an optimal refractive index range for efficient optical delivery. Accounting for three-dimensional angular effects and partial coupling losses, a refractive index of approximately 1.8–1.9 is found to be optimal. Based on this analysis,
Ohara S-TIH6
glass ($n = 1.8$) was selected as the waveguide material for the multi-blade design.

  
\begin{figure*}[htbp]
  \centering
  \includegraphics[width=0.8\textwidth]{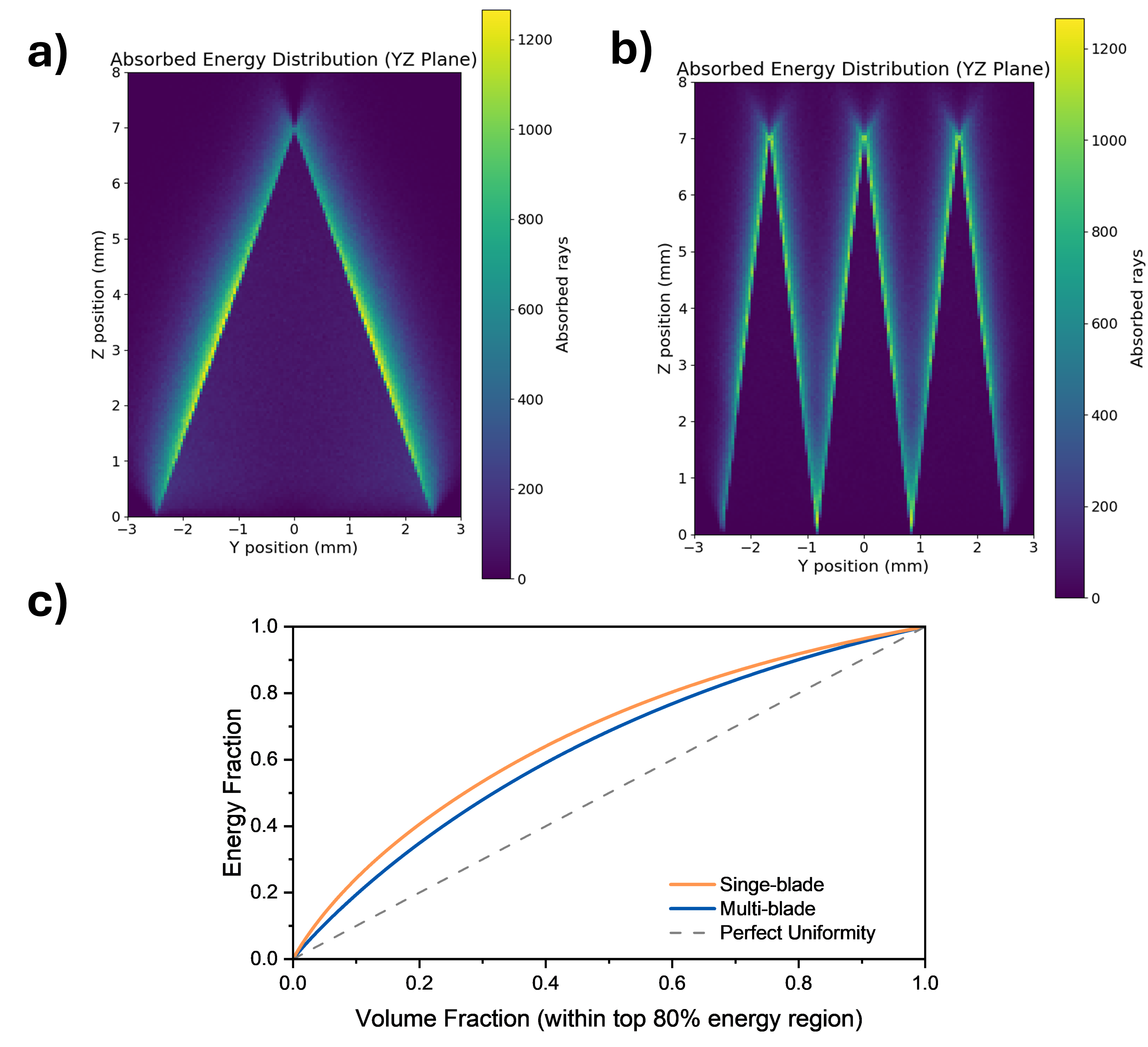}
  \caption{\textbf{Ray-tracing simulation results of different waveguide designs in the
  low-optical-power regime.} (a) Single-blade waveguide with refractive index of 1.45. (b) Multi-blade waveguide with refractive index of 1.8. The heat maps show the spatial distribution of absorbed photons within the Pc:PTP crystal for each design. (c) The curves show the cumulative fraction of absorbed energy as a function of the occupied crystal volume (voxel fraction) for the subset of voxels accounting for the top 80\% of the incident optical energy. A curve closer to the diagonal indicates more uniform absorption.
}
  \label{waveguide_simulation}
\end{figure*}

Due to the irregular geometry of the blade section, neither the angular nor the spatial distribution of rays can be treated analytically. Three-dimensional ray-tracing simulations were therefore performed using a modified version of the PVtrace software \cite{Farrell2008, FarrellPVTraceOriginal, verma2023ray, PVTraceModified}. Several kernel-level corrections were implemented to ensure physically consistent photon transport in embedded waveguide systems, including robust container identification, corrected Beer--Lambert path-length accounting, and proper enforcement of Fresnel boundary conditions. Details of these modifications are provided in Appendix III.

In all ray-tracing simulations, optical absorption within the Pc:PTP (0.1\% pentacene concentration) crystal is treated using a linear Beer--Lambert model with a single effective absorption coefficient. This approach does not explicitly capture bleaching, saturation, or other non-linear material responses. Instead, it enables a controlled comparison of injector geometries under identical effective absorption conditions, isolating purely geometric effects on the spatial power redistribution.

Figure~\ref{waveguide_simulation} shows the simulated spatial distribution of absorbed optical energy for two representative injector geometries: (a) a single-blade waveguide with refractive index $n=1.45$, and (b) the proposed multi-blade waveguide with refractive index $n=1.8$. The heat maps represent the projected three-dimensional distribution of absorbed photons within the crystal volume.

Consistent with this engineering interpretation of uniformity, we quantify spatial power redistribution using two complementary statistical metrics derived from the three-dimensional absorption maps: the effective uniform-area index $\mathbf{U}$ and the coefficient of variation $\mathbf{CV}$. Both metrics are computed over the subset of voxels that together account for the top 80\% of the \emph{incident optical energy}. Specifically, voxels are ranked according to their absorbed energy contribution, and the minimal set whose cumulative absorbed energy equals 80\% of the total incident optical energy is selected for analysis.
This definition ensures that the analyzed voxel set corresponds to the dominant optical power pathways inside the crystal and provides a consistent basis for comparing different injector geometries, independent of variations in total absorption efficiency.
By contrast, defining the analysis region relative to the absorbed energy would introduce a geometry-dependent normalization and obscure differences arising from optical redistribution and early exit losses.

The effective-uniform-area index \textbf{U} is defined as:
\begin{equation}
U = \frac{(\sum_{i=1}^{N} E_i)^2}{N \sum_{i=1}^{N} E_i^2}
\end{equation}
where $E_i$ is the absorbed energy in each voxel, and $N$ is the total number of voxels. This metric quantifies how effectively the total absorbed energy is distributed over the available volume: $\textbf{U}=1$ corresponds to perfectly uniform absorption, while smaller values indicate increasing spatial concentration of energy.
In contrast, the coefficient of variation \textbf{CV} is defined as:
\begin{equation}
CV = \frac{\sigma_E}{\mu_E}
\end{equation}
where $\sigma_E$ is the standard deviation of absorbed energy across all voxels, and $\mu_E$ is the mean absorbed energy. \textbf{CV} measures the relative spread of the absorption distribution and is particularly sensitive to local fluctuations.

While $\mathbf{U}$ and $\mathbf{CV}$ do not model bleaching or maser gain directly, they serve as proxies for spatial power concentration. High spatial concentration necessarily precedes any power-density–driven non-idealities, including local saturation, thermal effects, or bleaching. As such, these metrics provide a meaningful geometric comparison independent of the specific non-linear mechanism.

Two absorption conditions are considered. The first corresponds to the low optical power regime relevant to the present experiments (see Figure~\ref{waveguide_simulation}), where the effective absorption length is short and Beer--Lambert attenuation is strong. In this regime, the total absorbed energy is comparable across geometries, and the primary distinction lies in how that energy is spatially distributed. The corresponding uniformity metrics are summarized in Tables~\ref{tab:uniformity_metrics}.

The second condition represents a reduced average absorption strength, implemented by doubling the effective penetration depth. This scenario does not explicitly model bleaching dynamics; rather, it provides a geometric stress test that probes how injector designs perform when absorption becomes less effective.

This approximation is well justified for the multi-blade geometry, for which a substantial fraction of the output facet exhibits near-uniform intensity. Importantly, the same approximation is conservative for non-uniform geometries such as the single-blade design, as it neglects the preferential saturation and early loss expected in localized high-intensity regions.

Even under this favourable assumption, the single-blade geometry continues to exhibit pronounced spatial power concentration, whereas the multi-blade geometry maintains robust volumetric filling. The corresponding uniformity metrics are summarized in Tables~\ref{tab:uniformity_metrics_2}.

\begin{table}[]
    \caption{\textbf{Absorption uniformity metrics for different waveguide designs (low optical power regime).}}
    \label{tab:uniformity_metrics}
    \centering
    \begin{tabular}{lcc}
        \hline
        \textbf{Waveguide Design} & \textbf{U} & \textbf{CV} \\
        \hline
        Single-Blade & 0.635 & 0.759 \\
        Multi-Blade & 0.825 & 0.461 \\
        \hline
    \end{tabular}
\end{table}

\begin{table}[]
    \caption{\textbf{Absorption uniformity metrics for different waveguide designs (at reduced average absorption strength).}}
    \label{tab:uniformity_metrics_2}
    \centering
    \begin{tabular}{lcc}
        \hline
        \textbf{Waveguide Design} & \textbf{U} & \textbf{CV} \\
        \hline
        Single-Blade & 0.680 & 0.686 \\
        Multi-Blade & 0.836 & 0.444 \\
        \hline
    \end{tabular}
\end{table}

In the low optical power regime, the single-blade waveguide exhibits poorer baseline spatial uniformity than the multi-blade design, as reflected by its lower \textbf{U} and higher \textbf{CV}. This behavior originates from the strongly non-uniform angular (Lambertian) distribution of incident rays, which leads to pronounced spatial concentration of absorption within the crystal.

The multi-blade geometry mitigates this effect by redistributing incident optical power across multiple propagation pathways, resulting in a more homogeneous baseline absorption profile. These results establish the geometric baseline against which the injector performance under reduced absorption strength is evaluated.

When evaluated under reduced average absorption strength, the uniformity metrics reveal not only differences in spatial distribution, but also differences in sensitivity to changes in absorption. For geometries with strongly localized power delivery, reduced absorption is expected to amplify spatial non-uniformity, as the effective absorption volume expands preferentially into weakly illuminated regions.
This sensitivity is observed for the single-blade geometry, which continues to exhibit pronounced spatial power concentration under reduced absorption strength.
By contrast, the multi-blade geometry maintains a nearly unchanged level of uniformity, indicating a reduced sensitivity to variations in absorption strength. This robustness arises from its multi-path optical delivery, which promotes overlapping absorption regions and preserves volumetric filling even as the effective penetration depth increases.
The reduced sensitivity of the multi-blade design to absorption strength directly reflects its scalability advantage, as it remains effective under conditions where conventional non-uniform geometries become increasingly sensitive to optical attenuation.

To complement the spatial power distribution analysis, we evaluate the total fraction of incident optical power absorbed within the crystal volume for each injector geometry. This quantity provides a direct measure of geometric losses arising from premature exit or under-utilization of the crystal volume under identical absorption conditions.

In the low optical power regime, both injector designs achieve high absorption efficiencies, with 93.4\% for the single-blade and 98.5\% for the multi-blade geometry. This indicates that, under strong Beer--Lambert attenuation, geometric differences primarily affect how the incident optical power is spatially distributed rather than the overall absorption efficiency.

Under reduced average absorption strength, the total absorbed fraction decreases for both designs, reflecting increased geometric losses due to weaker attenuation. The reduction is significantly more pronounced for the single-blade geometry (80.1\%) than for the multi-blade design (91.7\%), indicating a higher sensitivity of the single-blade injector to reduced absorption.

This behavior is consistent with the stronger spatial power concentration observed in the single-blade case, which leads to a larger fraction of incident optical power exiting the crystal without being absorbed. These differences arise purely from geometric light redistribution under identical effective absorption conditions and do not rely on any explicit modelling of bleaching or other non-linear material responses.

It is important to note that this estimate is conservative for strongly non-uniform geometries such as the single-blade design. By enforcing a spatially uniform effective absorption coefficient, the present model does not capture the preferential saturation and early loss expected along localized high-intensity pathways. Consequently, the true geometric loss for single-path injectors under high-power conditions is expected to exceed the values predicted here.

By contrast, the reduced sensitivity of the multi-blade geometry to absorption strength reflects an intrinsic scalability advantage: its multi-path optical delivery suppresses localized over-saturation and preserves volumetric power utilization as pumping conditions are intensified.





\section*{Conclusions}
We have introduced and demonstrated a multi-blade invasive optical injector for LED-pumped room-temperature masers. By combining microscopic absorption modelling with three-dimensional ray-tracing, we decouple material nonlinearity from geometric light delivery and identify the pumping regimes in which injector geometry dominates performance. Our results show that the key advantage of the multi-blade injector lies not in idealized optical uniformity, but in its reduced sensitivity to absorption strength and its ability to suppress spatial power concentration. This geometric robustness enables scalable optical pumping under conditions where conventional single-path geometries become increasingly inefficient. The demonstrated injector–crystal assembly confirms the practical viability of this approach and establishes multi-path invasive pumping as a promising design strategy for future high-power, room-temperature maser systems.





\renewcommand\thesection{APPENDIX~\Roman{section}} 
\setcounter{section}{0}

\renewcommand\thefigure{S\arabic{figure}} 
\setcounter{figure}{0} 

\renewcommand\thetable{S\arabic{table}} 
\setcounter{table}{0} 

\begin{figure}[]
  \centering
  \includegraphics[width = 0.7\columnwidth]{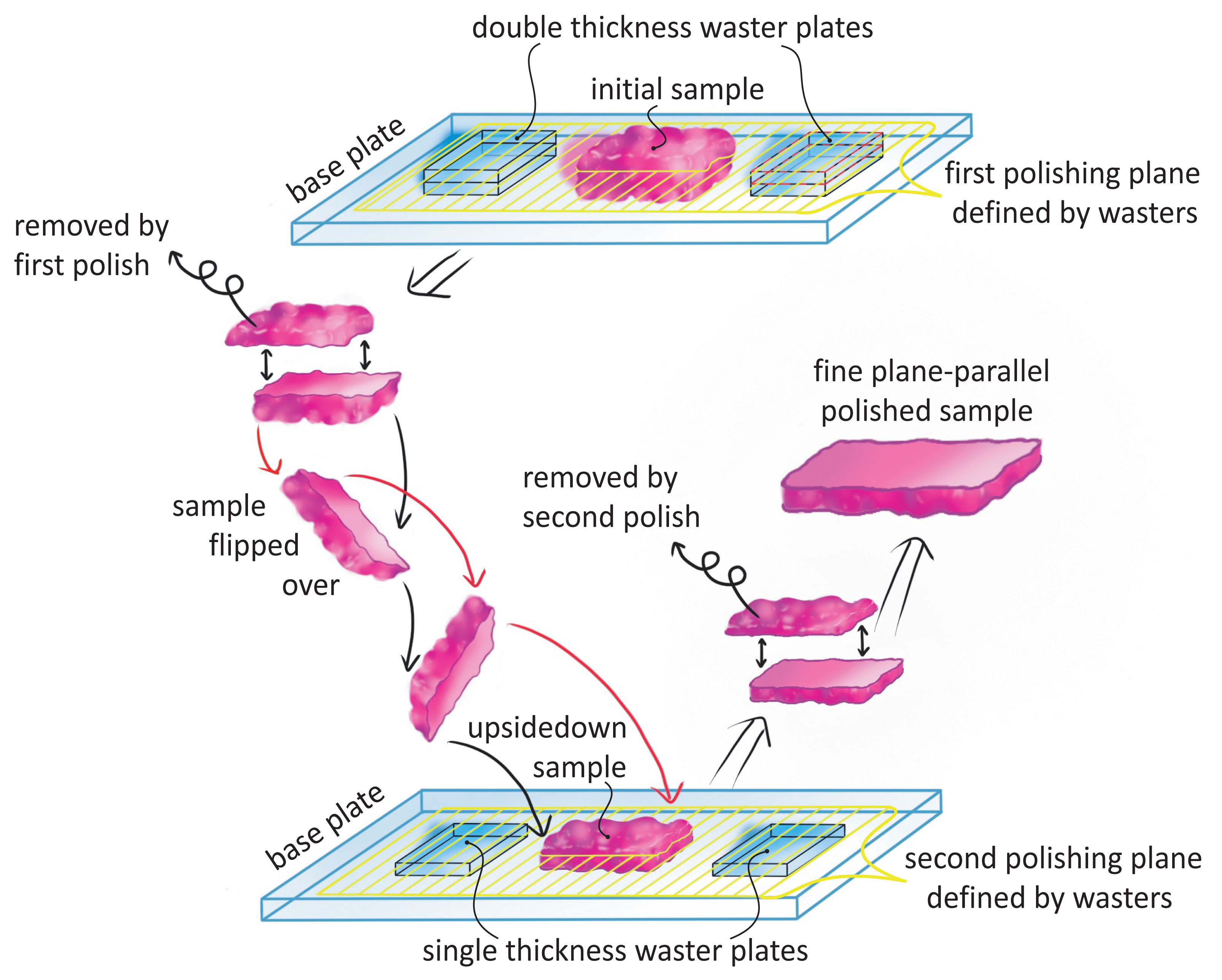}
  \caption{\textbf{Schematic illustration of the double-sided polishing strategy.} The Pc:PTP sample (pink) and reference block were first placed on the upper base plate and brought into contact with the grinding plane. To relieve asymmetric stress and suppress edge chipping during polishing, a stack of sacrificial “waster plates” (red dashed boxes) was positioned on the opposite side to balance the load. The entire assembly was then flipped vertically so that polishing proceeds with the previously lower face now facing the grinding plane. This alternating configuration ensures more uniform material removal and improves the overall flatness and integrity of the crystal.}
  \label{two_stage_pol}
\end{figure}

\textbf{\section{Experimental Details} }\label{sec:exp}

\textbf{Materials and Optical Sample Preparation}:
PTP was zone refined 20 times to achieve maximum purity ($\geq$99.99\%), and Pc was sublimated prior to growth. A crystal-growth tube with an inner diameter of 5.9 mm was filled with 0.1\% Pc:PTP powder and melted using the Bridgman method in a homemade furnace. 

Small, reasonably flawless single crystals at each doping concentration were first selected. Each selected crystal was then polished on both sides by first attaching it to a glass slide with optical wax on a hotplate then polishing the crystal’s top surface. The crystal was then dewaxed, flipped over onto its other side, re-waxed down and polished again. Note that the wax has two purposes: one is to secure the crystal so that it does not drop off from the glass slide while polishing; the other is to fill gaps compensating for the initial irregular shape of each crystal sample. 

“Waster plates” made of glass of known thickness (and much harder than the PTP being polished) were also attached with wax to the glass slides to control the crystal sample’s thickness; see Figure \ref{two_stage_pol}. After these steps, the crystals were dewaxed, cleaned by soaking in isopropanol and placed in labelled sample boxes. Note that our two-sided polishing method provides each optical sample with a “window” to look through the crystal, where 
these two polished surfaces making up the window are flat and parallel to each other. This is important to avoid bending light rays that passes through the sample in transmission.


\textbf{\section{1D Model for Optical Pumping Regime Identification} }\label{sec:sim_spin}

\subsection*{A.1 Four-level versus five-level model}

To validate the quasi-steady-state approximation applied to the $S_1$ state, the four-level model used in the main text was compared against a full five-level model across a wide range of pump intensities, including intensities exceeding those used experimentally by several orders of magnitude. In all regimes relevant to the present work, the four-level model reproduces the longitudinal absorption behaviour and triplet populations of the five-level model, while enabling substantially larger time steps and improved numerical stability.

\subsection*{A.2 Numerical stability and computational considerations}
In the full five-level formulation, the population of the $S_1$ state remains extremely small, making it susceptible to numerical instability unless prohibitively small time steps are used. The quasi-steady-state approximation eliminates this issue and allows the system to reach steady state within a feasible computational time, without affecting the predicted absorption behaviour in the regimes of interest.

\subsection*{A.3 Comparison with Takeda \textit{et al.}}

The model was further benchmarked against the laser-pumping regime studied by Takeda \textit{et al.}~\cite{takeda2002zero}, which operates on nanosecond timescales where spin-lattice relaxation and triplet sub-level dynamics are negligible. In this regime, Takeda’s three-level model, which treats the triplet manifold as a single effective state, provides an accurate description of absorption dynamics.

For moderate pump intensities, both the four-level model employed in the main text (using the quasi-steady-state approximation for $S_1$) and the full five-level model reproduce the similar absorption depth and profile predicted by Takeda’s treatment. At very high pump intensities, the quasi-steady-state approximation for $S_1$ breaks down, leading to huge deviations in the four-level model. Importantly, the full five-level model remains in good agreement with Takeda’s results in this regime, confirming that the observed discrepancies arise from the breakdown of the QSSA rather than from limitations of the underlying optical--spin-dynamics formulation.

\section{3D Ray-Tracing Simulation Details and Kernel Corrections}
\label{sec:sim}

This appendix documents the modifications made to the PVtrace ray-tracing framework to ensure physically consistent optical transport and absorption modelling in embedded waveguide systems. These technical details are provided to support numerical correctness and reproducibility and are not essential for following the main results.

\subsection*{A1. Robust container identification for multi-part geometries}

The standard \texttt{find\_container()} routine in PVtrace was revised to robustly handle small mesh overlaps arising from STL tessellation and numerical tolerances. In complex multi-part geometries, such overlaps can cause rays to incorrectly transition between adjacent solids, leading to non-physical propagation artifacts.

To address this issue, we implemented part-wise scene graph nodes with explicit container assignment and tolerance-aware intersection handling. This modification prevents spurious cross-part propagation and ensures that rays remain associated with the correct physical component throughout their trajectories.

\subsection*{A2. Absorption path integration and correction of spurious in-waveguide absorption}

In the unmodified PVtrace implementation, Beer--Lambert absorption sampling can incorrectly assign absorption events within the waveguide material when rays propagate at oblique angles or traverse multiple interfaces. This leads to artificial ``premature absorption'' and violates energy conservation.

To correct this behavior, absorption sampling was reformulated using a cumulative, direction-projected optical path length, weighted by $|\cos\theta|$, where $\theta$ is the local propagation angle relative to the waveguide axis. If a sampled absorption event falls within the waveguide, the interface interaction is first resolved and absorption updates are deferred until the ray enters the absorbing medium. This ensures that absorption is accumulated exclusively within the Pc:PTP crystal and that the sampled optical depth accurately reflects the true three-dimensional trajectory.

\subsection*{A3. Interface handling and Fresnel boundary enforcement}

To further ensure physically consistent transport, Fresnel reflection and transmission were enforced at all waveguide--absorber interfaces prior to any absorption updates. This guarantees that interface losses and angular redistribution are treated correctly and that absorption is not spuriously applied before boundary interactions are resolved.

\subsection*{A4. Additional simulation features and reproducibility}

Several auxiliary features were implemented to support the analyses presented in the main text, including multi-part STL geometry handling, planar detector nodes for directional flux readout, custom angular emission sampling based on measured LED distributions, and in-simulation recording of spatial absorption maps. Automatic kernel state resets were employed to ensure numerical reproducibility between simulation runs.

These modifications jointly provide geometrical robustness for multi-part scenes and physically consistent absorption in embedded waveguide--absorber architectures, which is critical to modeling maser-pumping light delivery.

\begin{figure}[]
  \centering
  \includegraphics[width = 0.9\textwidth]{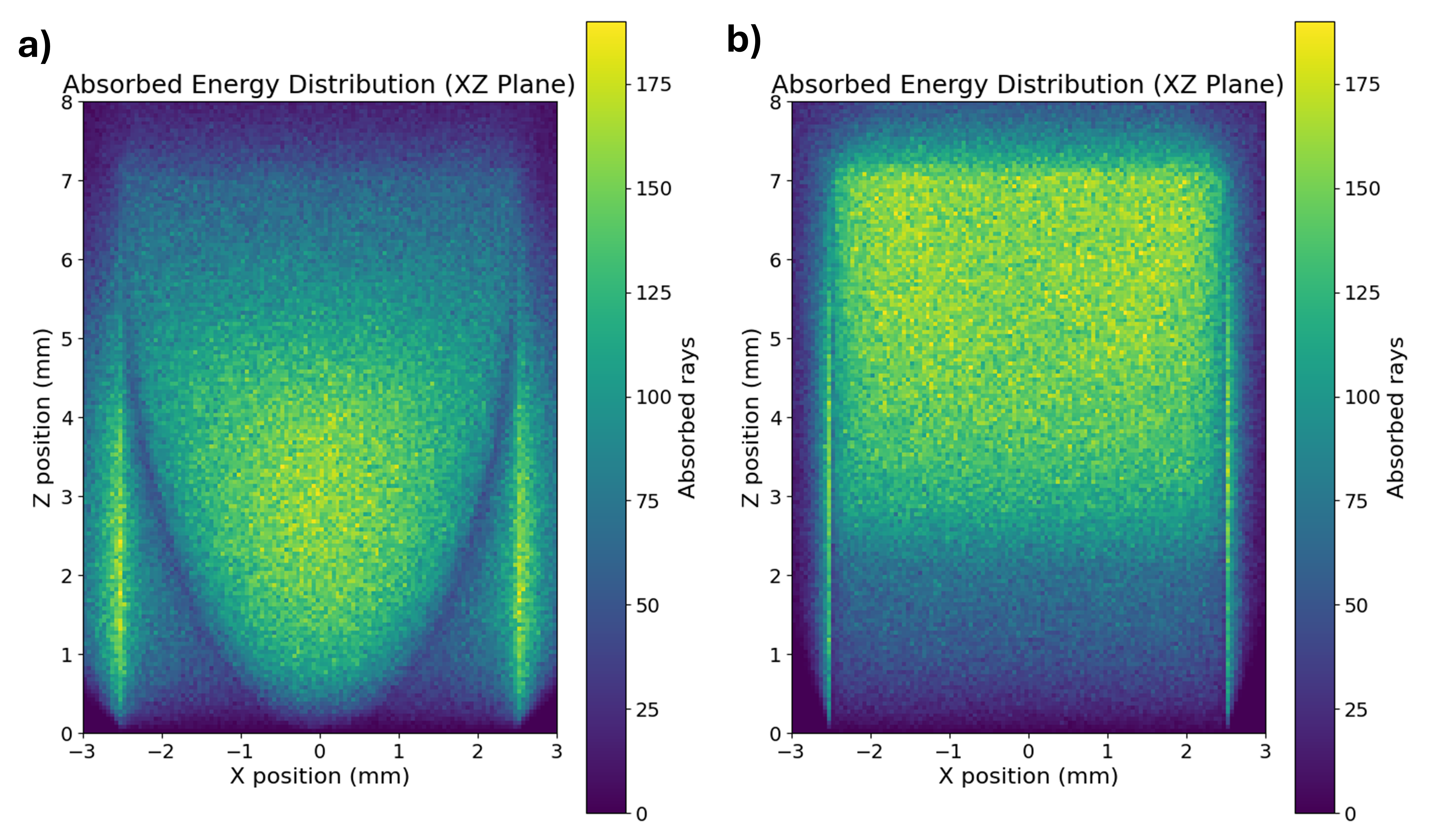}
  \caption{\textbf{Output-Facet Intensity Distribution of a): single blade design. b): multi-blade design} }
  \label{out_face}
\end{figure}

\subsection*{B. Scope and limitations}

We emphasize that the present analysis does not attempt to model bleaching, saturation, or other non-linear material responses at the output facet. The purpose of this appendix is solely to demonstrate that, from a geometric standpoint, the multi-blade injector produces an output intensity distribution for which a global effective absorption coefficient constitutes a reasonable and conservative approximation within the ray-tracing framework.

%




\vspace{1 cm}
\bibliography{references} 
\end{document}